\documentclass{article}
\usepackage{amsmath}
\usepackage{graphicx}
\graphicspath{{images/}}
\title{Note: Using Stein's estimator to correct the bound on the entropic uncertainty principle for more than two measurements}
\date{\vspace{-5ex}}
\author{Mark Stander \\ Electronic address: mark.stander@gmail.com}

\begin{document} 
\maketitle
\begin{abstract}
This note shows how to apply the James-Stein estimator to the case of entropic uncertainty relations of more than two observables. A better result is found compared to applying the ordinary estimator and we find a more optimal model compared with experimental data. The theoretical bounds for entropic relations for more than two observables are shown to be be improved.
\end{abstract}
\section{Background}
Uncertainty relations have been of interest in quantum mechanics since Heisenberg first proposed them in a thought experiment in the early 20th Century \cite{Heisenberg}. Later theoretical developments put uncertainty in terms of entropy \cite{Massen} and recently there has been a great deal of activity in looking at extending the relations to include more than two observables \cite{Winter}. More recently still, experimental work has validated these multi-observable predictions \cite{arxiv} \cite{Ma}.
Multi-observable entropic uncertainty relations are generally in the form:
\begin{equation}
    \sum_{k=1}^n H ( M_k )\leq B(M_1, M_2, M_3, ..., M_N, \rho),
\end{equation}
where $H$ is a measure of entropy, $M_m$ is a set of quantum measurements, and $B(\cdot)$ is some non-negative bound based on the measurements and density operator $\rho$ of the measured system. 

One such measure of uncertainty, and the measure used in the analysis provided below, is the special case of the R\'{e}nyi entropy, the Shannon entropy. The Shannon entropy is defined as:

\begin{equation}
  H(X) = \sum_{i=1}^n P(x_i) \log_2 P(x_i),
\end{equation}
where, in this context, $P(x_i)$ is the probability of measuring a quantum system in a particular state. In the analysis below, we measure the Shannon entropy of a measurement on a prepared quantum system by looking at the probabilities of finding that system in each of the possible states.

An example Shannon entropy of single measurement on a prepared quantum system, taken from the seminal work on finding an experimental basis for multi-measurement theory by Xing et al. \cite{arxiv}, can be seen below:

\begin{center}
 \begin{tabular}{||c c||} 
 \hline
 Probability of measuring system in state, initial state $|0>$ & \\ 
 \hline\hline
  $P(\sqrt{1/2}$, 0, $\sqrt{1/2})$ & 0.5496 \\
 \hline
 $P(\sqrt{1/2}$, 0, $-\sqrt{1/2})$ & 0.446 \\
 \hline
 $P(0, 1, 0)$ & 0.0044 \\
 \hline\hline
 Shannon entropy of single measurement of system, H & 1.0286 \\ [1ex] 
 \hline
\end{tabular}
\end{center}

The James-Stein estimator is one of the most fantastic results in statistical-mathematics. The James-Stein estimator shows that simultaneous estimation of three or more independent parameters is inadmissible using the standard, least-squares, estimator\cite{james1961estimation}.
In the case of the entropies measurements made on a system, the least squares estimator is:
\begin{equation}
    \hat{\theta}_{LS}=\mathbf{y}\ \ \  \text{where} \ \ \ \mathbf{y}=\{ H(M_1),H(M_2),...,H(M_n)\}
\end{equation}
where $H$ is the entropy of a measurement, where $\mathbf{y}$ is a vector of the entropies associated with each measurement, and $\hat{\theta}_{LS}$ is the least squares estimator. In this case, the best estimate of a parameter is the parameter itself. From the example from the table above, the least squares estimation of the entropy parameter would be $1.0286$. However, when simultaneous estimation of three or more independent parameters is made, as in the case of looking at the entropies of multi-measurement discussed below, the best estimate of the parameters involves James-Stein estimation. In other words, when the above vector $y$ is three or more dimensions, an alternative estimator must be used.

\maketitle
\section{Results}
To find the total entropy of multi-measurement you must sum the associated entropies of the individual measurements made. The literature thus far has assumed the least squares estimator dealt with above.
In terms of the least squares estimator, we would say:
Thus:
\begin{equation}
    \sum_{k=1}^n H ( M_k )=\mathbf{\sum y_k}
\end{equation}
However, Stein has shown the least squares estimator is inadmissible when dealing with the average of statistics across more than 3 measurements. The James-Stein estimator applies a factor to the measurements which, paradoxically, reduces the total Bayes risk across all of the measurements.
Let $\mathbf{y}$ be a vector of observations of length, $n$. In this case, $\mathbf{y}$ would be made up of the entropies associated with a measurement on the same initial state.
A better estimate of the true value of the vector $\mathbf{y}$ would be found by applying the James-Stein factor to the entropy associated with each of the measurements. The improved estimator would be
\begin{equation}
    \hat{\theta}_{JS}=(1-\frac{(n-2)\sigma^2}{\|y\|^2}\mathbf)\textbf{y}
\end{equation}
After applying the above factor, the measured entropies would be shrunk towards zero with the remarkable result that the sum of the entropies would better correspond with reality.

\section{Analysis}
In this section we offer an analysis of the above theory as applied to experimental multi-measurement data.
\begin{figure}[h]
\includegraphics[width=\textwidth]{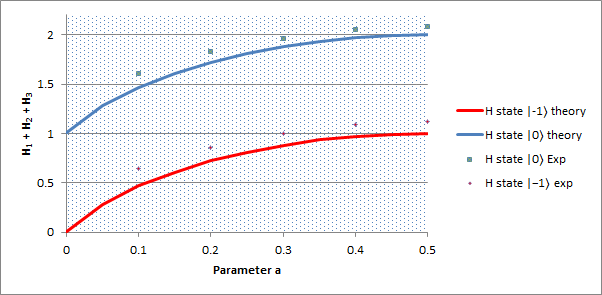}
\caption{Chart showing experimental findings vs. theoretical prediction}
\end{figure}
Chart reproduced from data collected from the ground-breaking work in entropic multi-measurement by Xing et al. \cite{arxiv}. The experimental work done compared the theoretical and experimental sums of Shannon entropies across three measurements on the same initial states. Two initial states are shown. The authors attributed the difference between experiment and theoretical model to the decoherence effect.

The theoretical predictions in these case are:
\begin{equation}
a \log_2(a) - (1-a)\log_2(1-a)
\end{equation} 
for the $|-1>$ initial state and 
\begin{equation}
a \log_2(a) - (1-a)\log_2(1-a) + 1
\end{equation} 
for the $|0>$ initial state, where $0 < a < 1$ is a parameter to be varied.

The James-Stein shrinkage of the entropies goes a very small way to explaining this delta.
\begin{figure}[h]
\includegraphics{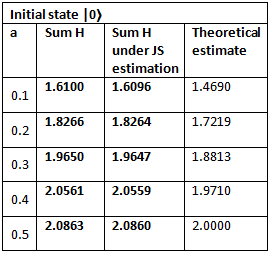}
\caption{Table showing comparison between experiment data, James-Stein shrunk data and theory for initial state \textbar 0\textgreater}
\end{figure}
\begin{figure}[h]
\includegraphics{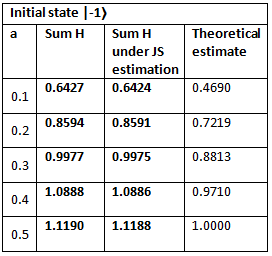}
\caption{Table showing comparison between experiment data, James-Stein shrunk data and theory for initial state \textbar -1\textgreater}
\end{figure}
We can see from Figure 2 and Figure 3 that applying the James-Stein estimator to the above measurements shrinks the experimental result towards the theoretical prediction in a small but significant way.
Alternatively, we might want to build the James-Stein estimator into our theoretical models in which case we would say that under James-Stein estimation the theoretical prediction has been stretched towards the experimental findings.

\section{Discussion}

We have shown that the theoretical prediction for entropic uncertainty relations with multi-measurements is improved by taking into account the James-Stein estimation. It is of note that the entropy across multi measurement is best estimated when the entropy of each individual measurement takes into account the entropies of the other measurements in the multi-measurement relationship. As with all James-Stein estimation, it should be taken into consideration that the entropy of each measurement is not necessarily improved by the shrinkage factor but the sum of those entropies necessarily is.

Appreciating the James-Stein effect also means that the theoretic entropy lower bounds could be increased to accommodate this new understanding.  For example, the multi-measurement theoretical lower bounds found by Liu et al. should be increased by the James-Stein shrinkage factor.

Liu et al. found a generalized bound on the entropic uncertainty relations for more than two measurements. For n observables, Liu et. al found that the state independent uncertainty relation \cite{liu2015entropic}:
\begin{equation}
    \sum_{k=1}^n H ( M_k ) \geq - \log b + (n - 1) S(\rho)
\end{equation}
Applying the James-Stein factor, the new equation for the total entropy of measurements across a state is:
\begin{equation}
    \sum _{k=1}^n \sum_{r=1}^n ( 1-\frac{(n-2)\sigma^2}{\sqrt{H(M_r)^2}})H(M_k)  \geq -\log b + (n - 1) S(\rho)
\end{equation}
In the above case, the standard estimator for the standard deviation should be used,
\begin{equation}
    \hat{\sigma}=\frac{1}{n}\sum(H(M_k)-\hat{H}(M_k))^2 ,
\end{equation}

the state of the measured system has its von Neumann entropy denoted by $S(\rho)$ and $n$ is the number of observables.

We also note an interesting extension will be the case where, from (5): 
\begin{equation}
    \frac{(n-2)\sigma^2} {||y||^2}  \to 1
\end{equation}
In this case, the estimate will tend to 0 and might offer a different view on the wavefunction collapse and measurement problem.

This note has shown an improved bound for entropic uncertainty relations for multiple measurements in terms of theory and experimental findings and shows an interesting new use case for the James-Stein estimator in physical theory.

\section{Acknowledgements}

I would like to thank Xing-Yu Pan and Heng Fan for kindly providing the data for the analysis. I would also like to thank Joy Christian and Seb Boissier for help navigating the world of academia.

\bibliographystyle{ieeetr}

\end{document}